\pdfoutput=1
\documentclass[10pt, conference, letterpaper]{IEEEtran}
\PassOptionsToPackage{hyphens}{url}
\IEEEoverridecommandlockouts
\usepackage{cite}
\usepackage{amsmath,amssymb,amsfonts}
\usepackage{amsthm}
\usepackage{graphicx}
\usepackage[dvipsnames,table,xcdraw]{xcolor}
\usepackage{colortbl}
\usepackage{subcaption}
\usepackage{algorithm}
\usepackage{algorithmic}
\usepackage{array}
\usepackage{caption}
\usepackage{booktabs}
\usepackage{xspace}
\usepackage{tcolorbox}
\usepackage{inconsolata}
\usepackage{hyperref}
\newcommand{\sys}{\textbf{\texttt{REBASE}}\xspace}

\theoremstyle{remark}
\newtheorem{observation}{Observation}

\colorlet{phaseI}{rgb:red!2,65;green!30,60;blue!20,125}
\colorlet{phaseII}{rgb:red!2,65;green!30,90;blue!20,125}
\colorlet{phaseIII}{rgb:red!60,100;green!20,90;blue!30,125}
\newcommand{\Comment}[1]{\hfill$\triangleright$ \textit{#1}}

\newcommand{\venue}[1]{{\scriptsize\color{gray}[#1]}}
\newcommand{\std}[1]{$_{\pm\text{\scriptsize #1}}$}
\definecolor{crystalblue}{rgb}{0.20,0.40,0.80}

\definecolor{darkgreen}{RGB}{0, 181, 18}
\newcommand{\gain}[1]{$\uparrow$ #1}
\newcommand{\mygreen}[1]{\cellcolor{darkgreen!#1}}

\definecolor{DeltaBg}{HTML}{D4F2D7}
\definecolor{SearchBg}{HTML}{C2E6F5}
\definecolor{MathBg}{HTML}{E6D4F2}
\definecolor{ScienceBg}{HTML}{FBE0BC}

\newcommand{\FreshGain}{45.6 points}
\newcommand{\StaleGap}{10.1 points}
\newcommand{\WalkShare}{62.2\%}
\newcommand{\DetectLag}{3}
\newcommand{\EvidenceRatio}{36.6$\times$}
\newcommand{\FixKnee}{79.1\%}
\newcommand{\KneeApps}{four of six}
\newcommand{\ScreenDelta}{9.4 points}
\newcommand{\DiffKB}{8.0\,KB}
\newcommand{\ShotKB}{298\,KB}
\newcommand{\QFingerprint}{27.4\%}
\newcommand{\QDiff}{79.1\%}
\newcommand{\ReplayCov}{86.7\%}

\newcommand{\RecoverGain}{56.1 points}
\newcommand{\RecoverGainCI}{[51.6, 60.6]}
\newcommand{\RecoverEp}{91.1\%}
\newcommand{\RecoverEpBlind}{68.9\%}
\newcommand{\ByteSave}{708$\times$}
\newcommand{\ByteSaveTraj}{215$\times$}
\newcommand{\CloudSave}{47.3$\times$}
\newcommand{\JouleSave}{2.5$\times$}
\newcommand{\JouleSaveRe}{2.3$\times$}
\newcommand{\AprimeGain}{3.7}
\newcommand{\StaleStepsH}{0.1}
\newcommand{\HMinusFresh}{0.4 points}
\newcommand{\HMinusFreshCI}{[$-$2.1, 3.0]}

\newcommand{\LinkRobustBw}{2.7 points}
\newcommand{\LinkRobustRtt}{4.0 points}
\newcommand{\FleetSlope}{80.0$\times$}
\newcommand{\Payback}{0.02}

\newcommand{\RecOneRe}{28.1\%}
\newcommand{\RecOneH}{83.7\%}
\newcommand{\RecOneCloudBlind}{76.3\%}
\newcommand{\JShot}{539\,J}
\newcommand{\JCloudOnly}{152\,J}
\newcommand{\JH}{702\,J}
\newcommand{\JSavedEpisode}{1028\,J}
\newcommand{\JEventMedian}{17\,J}
\newcommand{\XoneEvents}{135}
\newcommand{\XoneCloudRatio}{1.8$\times$}
\newcommand{\XoneStaleSteps}{2.4}
\newcommand{\XfourUpRatio}{3.2$\times$}
\newcommand{\XsixUpRatio}{14.4$\times$}
\newcommand{\DeltaXone}{3.3 points}
\newcommand{\DeltaXthree}{3.3 points}
\newcommand{\DeltaXseven}{9.8 points}
\newcommand{\DeltaXeight}{13.3 points}
\newcommand{\EventsH}{74}
\newcommand{\BeyondDiffPooled}{20.6\%}
\newcommand{\CloseByDiff}{79.4\%}
\newcommand{\BlindNoFailClock}{56.4\%}
\newcommand{\BlindH}{88.0\%}
\newcommand{\VisNoFailClock}{92.7\%}
\newcommand{\VisH}{96.7\%}
\newcommand{\FleetOverlap}{81.7\%}
\newcommand{\FleetD}{42}
\newcommand{\LatShotFast}{48\,s}
\newcommand{\LatShotSlow}{61\,s}
\newcommand{\LatHLo}{32.5\,s}
\newcommand{\LatHHi}{33.2\,s}

\providecommand{\RadarRelayoutLoss}{}\renewcommand{\RadarRelayoutLoss}{$-8.4$}
\providecommand{\RadarPolicyLoss}{}\renewcommand{\RadarPolicyLoss}{$-12.7$}

\newcommand{\ind}[1]{\mathbb{I}\!\left[\,#1\,\right]}
\newcommand{\Ed}{\mathcal{E}_{d}}
\newcommand{\Ec}{\mathcal{E}_{c}}
\newcommand{\Sk}[1]{\mathcal{S}_{#1}}
\newcommand{\Rop}{\mathcal{R}}
\graphicspath{{figures/}}
\hypersetup{pdfauthor={Beining Wu, Jun Huang, Yanxiao Zhao}, pdftitle={REBASE: Device-Cloud Experience Coherence for GUI Agents Across App Updates}, pdfkeywords={GUI agents, device-cloud systems, experience coherence, app updates, edge intelligence}}

\begin{document}
\bstctlcite{IEEEexample:BSTcontrol}

\title{REBASE: Device-Cloud Experience Coherence\\ for GUI Agents Across App Updates}

\author{\IEEEauthorblockN{Beining Wu\IEEEauthorrefmark{1}, Jun Huang\IEEEauthorrefmark{1}, and Yanxiao Zhao\IEEEauthorrefmark{2}}
\IEEEauthorblockA{\IEEEauthorrefmark{1}Department of Electrical Engineering and Computer Science, South Dakota State University, Brookings, SD 57007, USA\\
\IEEEauthorrefmark{2}Virginia Commonwealth University, Richmond, VA, USA\\
Email: Wu.Beining@jacks.sdstate.edu, Jun.Huang@sdstate.edu, yzhao7@vcu.edu}}

\maketitle

\begin{abstract}
A graphical user interface (GUI) agent that ships on a phone runs a small model and reuses experience: action paths recorded on earlier runs and cached from the cloud. When an app updates, part of this experience becomes silently wrong. On nine real app version pairs, an agent carrying experience recorded on the old version succeeds \StaleGap{} less often than one carrying none, and it does not notice the mismatch until \DetectLag{} steps after acting on it. We propose \sys{}, a protocol that keeps the device copy and the cloud copy of the experience coherent across updates. When a version changes, the cloud replays its copy on the new version and the device verifies each recorded step before executing it; when a step still fails, the device sends the cloud evidence in increasing size, starting from an accessibility-subtree diff, until the cloud re-derives an executable patch, keyed by version so that one repair serves the whole fleet. On two Jetson devices, \sys{} restores the success rate of stale experience to that of fresh experience recorded on the new version within one episode, at \ByteSave{} fewer bytes and \CloudSave{} fewer cloud calls than uploading screenshots to the cloud, and \JouleSave{} less device energy than running without experience.
\end{abstract}

\begin{IEEEkeywords}
GUI agents, device--cloud systems, experience coherence, app updates, edge intelligence
\end{IEEEkeywords}

\section{Introduction}
\label{sec:intro}

\IEEEPARstart{A}{} graphical user interface (GUI) agent on a phone runs a small model and reuses \emph{experience}: action paths recorded during earlier runs, consolidated by the vendor in the cloud, and cached on the device~\cite{Wu2026ARXIVForget,Wu2026ARXIVERRAND}. An app update can make these paths partly wrong without making them unusable: the failure is silent. Across nine real version pairs, an agent using experience from the previous version succeeds \StaleGap{} less often than one using none; in \WalkShare{} of executions, a step the update broke still finds an element to act on, so the agent does not notice the mismatch until \DetectLag{} steps later. Two signals go unused: the vendor's cloud can obtain the new version before the device does, and the device knows it has updated before it fails.

Device--cloud agent and inference systems decide which steps or model parts run in the cloud and what the device uploads~\cite{Zhou2025ARXIV,Jiang2026ACL,Li2026ARXIV,Fan2025NeurIPS,Xie2026INFOCOM,Wang2026INFOCOM,Zhang2026ARXIV,Fang2025JSAC}, and use the cloud to finish the current task, not to correct experience for later runs: MAI-UI uploads screenshot history on failure, and nothing returns to change the device's next run~\cite{Zhou2025ARXIV}. Experience and memory systems keep experience on the machine that uses it, treat the app as fixed, and discuss updates only as a limitation~\cite{Qin2026ICML,Mi2026ICML,Men2026ACL}. CoMIC uploads complete trajectories and filters them in the cloud~\cite{Wang2026ARXIVa}, whereas OpenJarvis improves the on-device agent in an offline search phase and ships the edits down~\cite{SaadFalcon2026ARXIV}. Self-healing test automation repairs a broken element locator offline on one machine~\cite{SelfHealing2026WEB,Cao2024ICSE,Cao2026PACMSE}, a relocation we adopt. There, a failed script stops; an agent instead continues along a stale path, and a second copy of its experience sits in the cloud. \emph{Which evidence should cross the device--cloud link after an update, and when, remains open.}

\sys{} (\underline{\textbf{Re}}play-\underline{\textbf{B}}acked \underline{\textbf{A}}ttestation with \underline{\textbf{S}}taged \underline{\textbf{E}}vidence) treats an app update as a \emph{coherence event}: the device copy and the cloud copy of the agent's experience stop agreeing. \sys{} reconciles them on two clocks. On the \emph{update clock}, which ticks when a version changes, the cloud replays its copy on the new version and the device verifies each step before executing it. On the \emph{failure clock}, which ticks when a step fails, the device sends evidence in increasing size, from a fingerprint and an accessibility-subtree diff up to a screenshot, until the cloud re-derives an executable patch. The link carries a description of what changed, and a screenshot only when smaller evidence fails. Our contributions are threefold:

\begin{itemize}
\item We measure how consolidated experience fails after an update and what repairs it: the failure is silent, and an accessibility-subtree diff lets the cloud repair \FixKnee{} of failing steps with a per-step median of \EvidenceRatio{} fewer bytes than a screenshot.
\item We design \sys{}, a two-clock protocol that selects evidence by the event that triggered it.
\item We key every repair by version, so cloud re-derivations grow with the steps an update breaks, not with the fleet: in a trace-driven fleet of 100 devices, \sys{} re-derives \FleetSlope{} fewer times than without it.
\end{itemize}

We implement \sys{} on two Jetson AGX Orin devices and a Thor gateway, with nine Android apps on two releases each. It restores stale experience to the success rate of fresh experience within one episode, at \ByteSave{} fewer bytes and \CloudSave{} fewer cloud calls than uploading screenshots and \JouleSave{} less device energy than running without experience. To our knowledge, \sys{} is the first system that keeps a GUI agent's experience coherent across app updates over a device--cloud link and measures its cost on real devices.

\section{Background and Observations}
\label{sec:obs}

\begin{figure*}[t]
\centering
\includegraphics[width=0.377\textwidth]{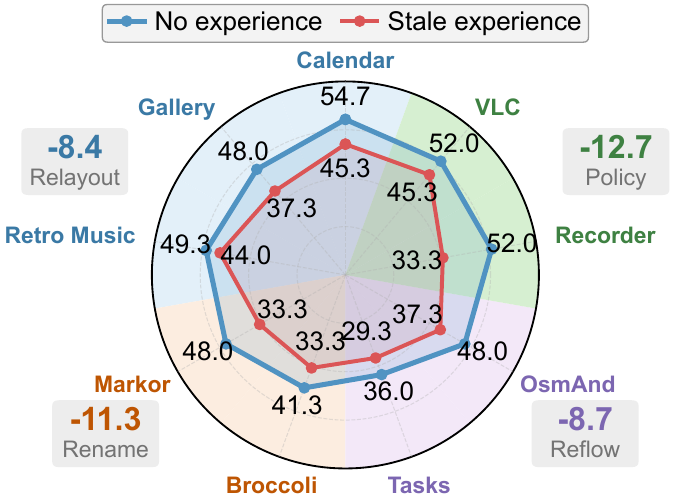}\hfill%
\includegraphics[width=0.595\textwidth]{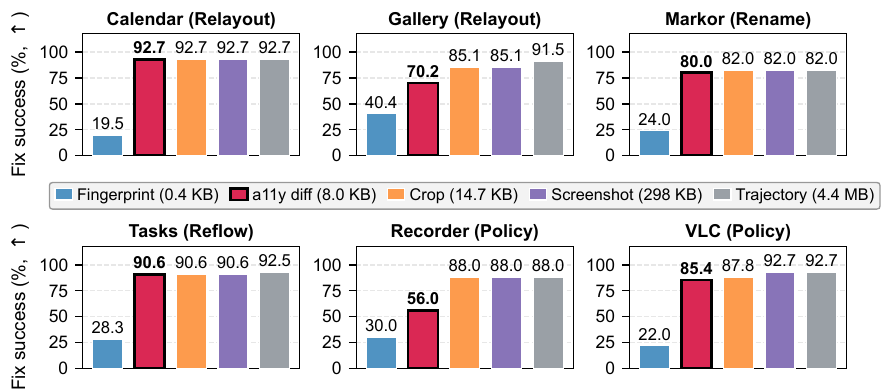}
\caption{\textbf{Stale experience fails silently, while accessibility-subtree diffs enable most repairs.} \textbf{Left:} Success on new versions of nine apps with no experience or old-version experience; badges report mean loss by update type. \textbf{Right:} Share of failing steps repaired from five evidence levels across six cloud-visible apps. The diff marks the knee in \KneeApps{} apps. Three seeds.}
\label{fig:teaser}
\end{figure*}

\subsection{Device and Cloud Copies of GUI Experience}
\label{sec:obs-setting}
A deployed GUI agent runs a small model on the device and reuses \emph{experience}~\cite{Wu2026TNSE,Wu2026ARXIVLifecycle}. For app version $v$, experience comprises executable paths bound to accessibility-tree elements~\cite{Android2026WEB}. Each step uses a \emph{fingerprint} of the element's role, label, and position. The agent \emph{walks} a path by executing its steps in order. The cloud consolidates these paths, and the device caches a copy~\cite{Wu2026ARXIVCrystalMem,Fang2025ARXIV}. We denote the device and cloud copies by $\Ed(v)$ and $\Ec(v)$, respectively. They agree until the app updates; either side may see the new version first, and a wireless link joins them~\cite{Wu2026TON,Dong2026TCCN,Xing2025ACR}. Experience recorded on $v$ but walked on the next version $v'$ is \emph{stale}~\cite{Wu2023ACCESS,Ding2026ICDCS}.

\subsection{Observation: Silent Failure of Stale Experience}
\label{sec:obs-1}
\begin{tcolorbox}
\begin{observation}
\label{obs:stale}
Stale experience lowers success by \StaleGap{} relative to no experience; update-broken steps remain walkable in \WalkShare{} of executions, delaying detection by \DetectLag{} steps.
\end{observation}
\end{tcolorbox}
Fig.~\ref{fig:teaser} (left) compares success on $v'$ with no experience and with $\Ed(v)$. The stale polygon lies inside the no-experience polygon on all nine axes. Relayouts move elements, renames change labels, reflows add or remove steps, and policy updates add a dialog or permission prompt. Mean loss by update type ranges from \RadarRelayoutLoss{} points for relayouts to \RadarPolicyLoss{} for policy updates. Fresh experience recorded on $v'$ raises success by \FreshGain{} over no experience; stale experience does not retain this gain~\cite{Wu2026ICDCS,Wu2026TMM}.

A broken step fails silently when it remains walkable. Its fingerprint still matches an element by role and label, but the action produces the wrong next screen. The agent continues until a later element is missing or the episode ends.

\subsection{Observation: Repair Value of Accessibility-Subtree Diffs}
\label{sec:obs-2}
\begin{tcolorbox}
\begin{observation}
\label{obs:diff}
An accessibility-subtree diff repairs \FixKnee{} of failing steps; a screenshot adds only \ScreenDelta{} at a per-step median of \EvidenceRatio{} the bytes.
\end{observation}
\end{tcolorbox}
Fig.~\ref{fig:teaser} (right) compares five evidence types on the same failing steps. A fingerprint repairs \QFingerprint{}, while the \DiffKB{} accessibility-subtree diff raises the rate to \QDiff{}. An element crop, the \ShotKB{} screenshot, and the complete trajectory each add about \ScreenDelta{}. The knee, the first level after which more evidence adds little, is the diff for \KneeApps{} apps and the crop for the other two.

An update changes only a few elements around the broken one. The diff isolates these changes and shows the recorded element in its new neighborhood, so the model can relocate it. For the same step, a screenshot needs a median of \EvidenceRatio{} the bytes of the diff but does not identify which elements changed. The trajectory adds execution history unused by relocation.

\noindent\textbf{Opportunities.}
Obs.~\ref{obs:stale} motivates the update clock: the cloud replays when it receives $v'$ first, and the updated device verifies each recorded step before execution. Obs.~\ref{obs:diff} motivates the failure clock: the device sends evidence of the change after a failure and escalates only when repair fails. \sys{} combines both triggers in one protocol.

\section{\sys{} Design}
\label{sec:design}

\begin{figure*}[t]
\centering
\includegraphics[width=0.95\textwidth]{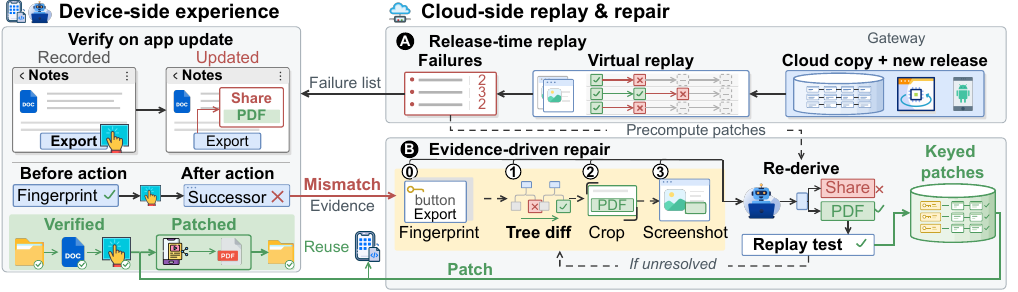}
\caption{\sys{} overview. The update clock drives verification on the device, which checks each recorded step before and after acting, and release-time replay on the cloud (A), which precomputes patches. The failure clock drives evidence-driven repair (B): the lightest sufficient evidence re-derives the failed segment, which is tested and returned under its version key for reuse.}
\label{fig:arch}
\end{figure*}

\subsection{Overview and Problem Formulation}
\label{sec:design-overview}

\sys{} has three agents~\cite{Wu2023MPE,Ding2026ICNC}. The \emph{device agent} runs the small model and fingerprint executor over $\Ed(v)$, whose paths were recorded for installed version $v$. The \emph{cloud agent} maintains $\Ec(v)$ and a strong model. The \emph{gateway}, on the cloud side, monitors releases, runs a virtual device, and indexes coherence events by version key. One protocol follows two clocks (Fig.~\ref{fig:arch}): the \emph{update clock} ticks at a version change, and the \emph{failure clock} at a verification mismatch on the device, which raises a \emph{coherence event}.

Formally, a path $p=\langle(f_1,a_1),\dots,(f_n,a_n)\rangle$ contains steps that apply action $a_k$ to the element identified by fingerprint $f_k=(r,t,\tilde b,h)$. Its fields are the role, text label, normalized box, and two-ancestor hash. Fingerprints match when
\begin{equation}
f\simeq f'\;\Longleftrightarrow\;(r,h)=(r',h')\ \wedge\ d(t,t')\le\theta_t\ \wedge\ \Delta(\tilde b,\tilde b')\le\theta_b.
\label{eq:match}
\end{equation}
Here, $d$ is normalized label edit distance, and $\Delta$ is center displacement in screen units; $\Delta=0$ when the boxes overlap by at least half. The default thresholds are $\theta_t=0.2$ and $\theta_b=0.1$. Let $\Sk{}$ denote a screen's fingerprint set. We write $f\simeq\Sk{}$ when $f$ matches an element of $\Sk{}$ and use the same relation for screens whose sets overlap by at least $\theta_s=0.5$. Verification $\nu_k$ is one when the step-$k$ element exists on $\Sk{k}(v')$ and the resulting screen matches the recorded successor $\Sk{k+1}^{\star}$. With $\mathbb{I}[\cdot]$ as the indicator, $D(v\!\to\!v')$ contains each path's first failing step:
\small
\begin{equation}
\begin{aligned}
\nu_k(p,v')&=\ind{f_k\simeq\Sk{k}(v')}\;\ind{\Sk{k+1}(v')\simeq\Sk{k+1}^{\star}},\\
D(v\!\to\!v')&=\bigl\{(p,k):\ \nu_k(p,v')=0,\ \nu_j(p,v')=1\ \ \forall j<k\bigr\}.
\end{aligned}
\label{eq:coh}
\end{equation}
\normalsize
A first-factor pass followed by a second-factor failure is the stale-but-walkable step of Obs.~\ref{obs:stale}. \sys{} selects the answering level $\ell_e$ of each event $e$ and an update-time push set $\mathcal{U}$ to maximize autonomous success over the first $K$ episodes~\cite{Huang2025RACS,Huang2025TMC,Wu2025WASA}:
\small
\begin{equation}
\begin{aligned}
\max_{\{\ell_e\},\,\mathcal{U}}\ \ \mathbb{E}\Bigl[\,\sum_{k=1}^{K}y_k\Bigr]\quad
\text{s.t.}\quad&\sum_{p\in\mathcal{U}}\upsilon_p\le\bar\upsilon,\\
&\mathbb{E}\Bigl[\sum_{e\in\mathcal{C}}\beta_e\Bigr]\le\bar\beta,\qquad
\mathbb{E}\Bigl[\sum_{e\in\mathcal{C}}\varepsilon_e\Bigr]\le\bar\varepsilon.
\end{aligned}
\label{eq:obj}
\end{equation}
\normalsize
Here, $y_k$ marks autonomous success in episode $k$ and $\upsilon_p$ the size of a pushed entry. For event $e$, $\beta_e$ and $\varepsilon_e$ denote bytes and joules; $\mathcal{C}$ is the event set, and barred terms are budgets. The recovery index $T_r$ is zero if all episodes succeed, $\max\{k:y_k=0\}$ if recovery occurs by $K$, and $K+1$ otherwise. Since $\sum_{k=1}^{K}y_k\ge K-T_r$, the objective favors $T_r\in\{0,1\}$. Equation~\eqref{eq:obj} selects link evidence; execution remains on the device and repair in the cloud, where the strong model runs, a candidate patch can be tested on the virtual device rather than on the user's app~\cite{Wu2026MNET,Ding2026ARXIVTwinLoop}, and a tested patch serves every device under its version key. When budget permits, $\mathcal{U}$ contains replayable $D$ entries; each push costs about as much as the smallest evidence and prevents a later event. A stale-but-walkable step executes at most once per device, because its successor screen is observed only after the action; every other executed step passes $\nu_k$ on $v'$ or comes from a patch committed for $v'$.

\subsection{Cloud Replay and On-Device Verification}
\label{sec:design-update}

When $v'$ is obtainable, the gateway installs it on the virtual device and replays each path in $\Ec(v)$ until $\nu_k$ fails, without a model call. The list $I(v')=\{(p,k)\in D(v\!\to\!v'):p\in\mathcal{V}\}$ restricts $D$ to the replayable set $\mathcal{V}$. Its share $\gamma=|\mathcal{V}|/|\Ec(v)|$ is \ReplayCov{} on our apps; paths blocked by login, payment, or device state remain for device verification. The gateway re-derives a patch for each entry, stores it in the patch set $P(v')$, pushes one bit per path plus a step index, and serves patches on demand.

If an update precedes the push, or the cloud could not replay a path, the device marks the app's paths \textsc{Unverified}$(v')$. During execution, it tests $\nu_k$ without a model call: the first factor before the action and the second afterward. A failure raises an event. A stale-but-walkable step therefore costs one action rather than the lag in Obs.~\ref{obs:stale}; passing steps form a \emph{verified prefix}. For a flagged step, the device pulls its patch before reaching it. Phases~I and~II of Algorithm~\ref{alg:rebase} cover both cases.

\subsection{Staged Evidence and Patch Return}
\label{sec:design-failure}

The failure clock sends the smallest evidence first~\cite{Fang2025TON} (Obs.~\ref{obs:diff}).

The re-derivation operator $\Rop$ maps evidence $\xi_\ell$ to a verified patch or $\bot$. Level $\xi_0$ contains the fingerprint and failing step; with app $\alpha$ and version $v'$, they identify the version key. Level $\xi_1$ is the recorded-to-current accessibility-subtree diff, rooted at the expected element's grandparent and extending three levels with one sibling hop. Levels $\xi_2$ and $\xi_3$ contain a low-resolution crop around the expected box and the screenshot, with bytes $\beta_0<\beta_1<\beta_2<\beta_3$. As in locator repair~\cite{Cao2024ICSE,SelfHealing2026WEB}, the cloud proposes candidate elements and tests the step on the virtual device. It commits a successful patch to $P(v')$ or returns $\bot$. The first successful level resolves the event. Expected bytes depend on conditional success $q_\ell=\Pr[\Rop(\xi_\ell)\neq\bot\mid\Rop(\xi_m)=\bot,\ m<\ell]$:
\small
\begin{equation}
\ell^{\star}_{e}=\min\bigl\{\ell:\ \Rop(\xi_\ell)\neq\bot\bigr\},\qquad
\mathbb{E}\bigl[\beta_e\bigr]=\textstyle\sum_{\ell}\beta_\ell\prod_{m<\ell}\bigl(1-q_m\bigr).
\label{eq:ladder}
\end{equation}
\normalsize
Past the knee, a crop or screenshot adds about \ScreenDelta{} at a multiple of the diff's bytes, so the ladder closes most events at $\xi_0$ or $\xi_1$. Because the ladder ends at the screenshot, it repairs every step that a screenshot alone repairs, at the expected bytes of~\eqref{eq:ladder} rather than $\beta_3$ per event. \sys{} starts at $\xi_0$, advances only after $\bot$, and returns to the small model if all levels fail. Table~\ref{tab:main} reports fixed-diff and fixed-screenshot alternatives. Patch $\pi=\langle(f'_j,a'_j)\rangle_{j=k}^{k+m}$ replaces the segment of $p$ from failing step $k$~\cite{Ding2026ARXIVEASE,Ding2026MNET}; the executor replays its actions so the small model can apply a cloud-derived repair (Phase~III of Algorithm~\ref{alg:rebase}).

\begin{algorithm}[t]
\caption{\sys{}: one protocol, two clocks
(\colorbox{phaseI}{Replay}, \colorbox{phaseII}{Verify}, and
\colorbox{phaseIII}{Escalate-and-Patch})}
\label{alg:rebase}
\textbf{Input:} device copy $\Ed(v)$, cloud copy $\Ec(v)$; ladder
$\xi_0,\dots,\xi_3$
\begin{algorithmic}[1]
\STATE \textit{Phase I: Replay} \Comment{cloud side, when $v'$ ships}
\colorbox{phaseI}{
\parbox{0.82\columnwidth}{
\STATE replay $\Ec(v)$ on the virtual device running $v'$
\STATE $I(v')\gets\{(p,k)\in D(v\!\to\!v'):\ p\in\mathcal{V}\}$ by \eqref{eq:coh}
\STATE re-derive and store in $P(v')$ one patch for each entry of $I(v')$
\STATE push $I(v')$; serve $P(v')$ on demand
}}

\STATE \textit{Phase II: Verify} \Comment{device, on package update}
\colorbox{phaseII}{
\parbox{0.82\columnwidth}{
\STATE mark the app's paths \textsc{Unverified}$(v')$
\FOR{each step $k$ of an executing path $p$}
\STATE \textbf{if} $\nu_k(p,v')=1$ by \eqref{eq:coh} \textbf{then} extend the prefix
\STATE \textbf{else} raise an event with $\xi_0$; \textbf{break}
\ENDFOR
}}

\STATE \textit{Phase III: Escalate-and-Patch} \Comment{on a coherence event}
\colorbox{phaseIII}{
\parbox{0.82\columnwidth}{
\STATE $\kappa\gets(\alpha,v',f_k)$ \Comment{version key}
\STATE \textbf{if} $\kappa\in P(v')$ \textbf{then} $\pi\gets\pi_\kappa$; go to line~\ref{ln:apply}
\FOR{$\ell=0,\dots,3$}
\STATE send $\xi_\ell$;\quad $\pi\gets\Rop(\xi_\ell)$ by \eqref{eq:ladder}
\STATE \textbf{if} $\pi\neq\bot$ \textbf{then break}
\ENDFOR
\STATE \textbf{if} $\pi=\bot$ \textbf{then return} to the small model
\STATE commit $\pi$ under $\kappa$;\ apply $\pi$ in place; continue\label{ln:apply}
}}
\end{algorithmic}
\end{algorithm}

\subsection{Version Keys and Fleet Deduplication}
\label{sec:design-fleet}

A release may break the same fingerprint on every device. \sys{} keys events by $\kappa=(\alpha,v',f)$, where $\alpha$ identifies the app, rather than by device. The first report of $\kappa$ triggers one re-derivation; later reports send only $\xi_0$ and receive $\pi_\kappa$ from $P(v')$~\cite{Tchalla2026ARXIV,Ding2025IPCCC,Wu2026IoTJ}. If $N_\kappa$ devices report $\kappa$, the cloud performs $M=\sum_{\kappa}\ind{N_\kappa\ge1}\le|D(v\!\to\!v')|$ re-derivations. Without the key, $\tilde M=\sum_{\kappa}N_\kappa\approx N\omega\,|D(v\!\to\!v')|$, where $\omega$ is the fraction of $N$ devices that break on that fingerprint. Keyed re-derivation therefore depends on distinct update-broken fingerprints, not fleet size.

\subsection{Coverage and Fallbacks}
\label{sec:design-scope}

If a screen lacks an accessibility tree, as with a canvas or opaque web view, $\xi_1$ is empty and the ladder moves from $\xi_0$ to $\xi_2$. Every measured failure in our nine apps exposed a subtree. Paths unreachable on the virtual device, and apps unavailable to the cloud, skip replay but retain device verification and the failure clock~\cite{Pan2023SCIS}. The device confirms each returned patch through the next verified step.

\section{Implementation and Testbed}
\label{sec:impl}

\begin{figure}[t]
\centering
\includegraphics[width=\columnwidth]{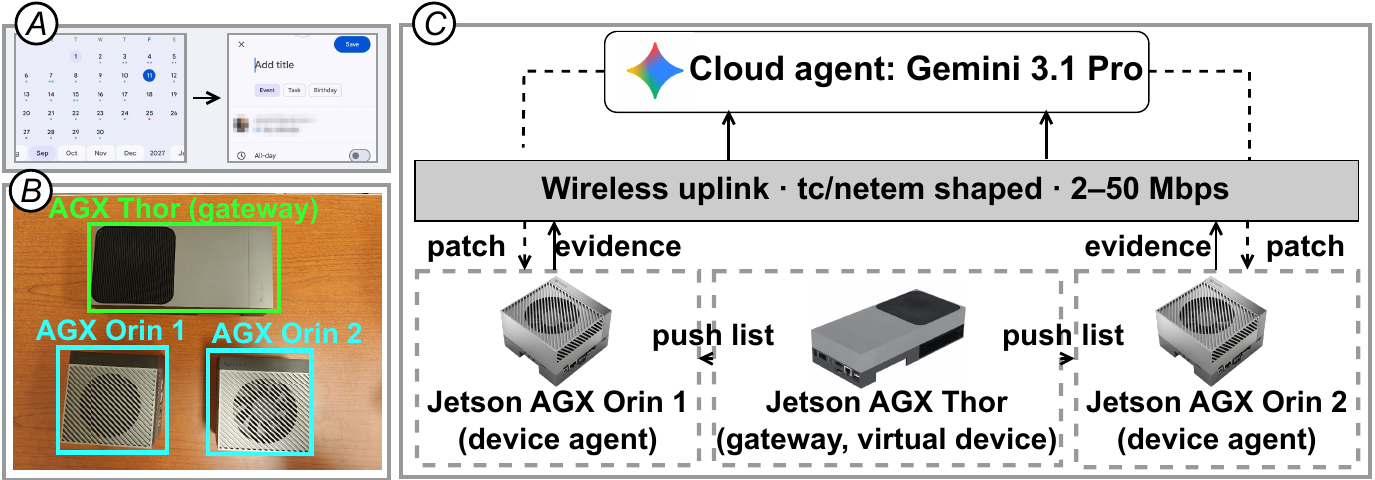}
\caption{The \sys{} testbed: (A) two screens of a walked path, (B) the devices, and (C) the topology.}
\label{fig:testbed}
\end{figure}

Two NVIDIA Jetson AGX Orin 64\,GB boards are the devices (Fig.~\ref{fig:testbed}); each runs Qwen3-VL-8B-Instruct~\cite{Bai2025ARXIVb} at INT4 with activation-aware weight quantization and the fingerprint executor. Both control Android~13 AndroidWorld environments~\cite{Rawles2025ICLR} hosted on a Jetson AGX Thor gateway. \texttt{tc/netem} shapes the gateway--cloud uplink to three bandwidth and three round-trip-time tiers; the cloud agent uses Gemini~3.1~Pro.

At each step, we record a quality-85 JPEG screenshot, accessibility forest, screen hash, evaluator verdict, bytes, and power trace on a common clock. Byte counts cover gzipped JSON and encoded images at the application layer. Device energy is the sum of three INA3221 rails sampled at 10\,Hz through tegrastats, minus session idle power.

The Thor runs a separate AndroidWorld image as the virtual device, hosts the release watcher and fleet index, and joins the two Orins as a third device in the fleet experiment. We test nine AndroidWorld apps; $v$ is the version AndroidWorld pins, and $v'$ is the next F-Droid Archive release that changes an accessibility tree along a recorded path. For three blind pairs, a self-hosted store supplies $v'$ but hides it from the cloud.

\section{Evaluation}
\label{sec:eval}

\begin{table*}[t]
\centering
\caption{\textbf{\sys{} recovers stale experience to fresh-experience success within one episode at a small fraction of the bytes and cloud calls of the takeover and upload baselines.} Main tier, $K=5$, three seeds. Family best \underline{underlined}; \colorbox{DeltaBg}{$\Delta$} = \sys{}'s Succ@$K$ gain (points; shade $\propto$ gain); \textcolor{gray}{\textit{gray italic}} = cloud-only bound, unranked; ablation rows: $\Delta$ to full \sys{}.}
\label{tab:main}
\footnotesize
\setlength{\tabcolsep}{2.8pt}
\renewcommand{\arraystretch}{1.12}
\begin{tabular}{l|cc|cc|cc|c}
\toprule
& \multicolumn{2}{c|}{\cellcolor{SearchBg}\textbf{Capability}}
& \multicolumn{2}{c|}{\cellcolor{MathBg}\textbf{Wire}}
& \multicolumn{2}{c|}{\cellcolor{ScienceBg}\textbf{Energy \& cloud}}
& \\
\cmidrule(lr){2-3}\cmidrule(lr){4-5}\cmidrule(lr){6-7}
\textbf{Method} & \textbf{Succ@$K$}$\,\uparrow$ & \textbf{$T_r$}$\,\downarrow$ & \textbf{Up KB}$\,\downarrow$ & \textbf{Down KB}$\,\downarrow$ & \textbf{Device J}$\,\downarrow$ & \textbf{Cloud calls}$\,\downarrow$ & \textbf{$\Delta$} \\
\midrule
{\color{gray}\textit{Cloud-only agent}} & {\color{gray}\textit{97.5\std{1.1}}} & {\color{gray}\textit{6}} & {\color{gray}\textit{3825.7}} & {\color{gray}\textit{3.76}} & {\color{gray}\textit{152}} & {\color{gray}\textit{12.57}} & {\color{gray}\textit{$-$3.7}} \\
\midrule
No experience & \underline{47.7\std{2.0}} & 6 & 0.0 & 0.00 & 1730 & 0.00 & \mygreen{69}\gain{46.1} \\
Stale experience, unrepaired & 37.6\std{2.5} & 6 & 0.0 & 0.00 & \underline{940} & 0.00 & \mygreen{84}\gain{56.1} \\
\midrule
Re-explore on failure & 69.3\std{4.5} & 3 & \underline{0.0} & \underline{0.00} & 1612 & \underline{0.00} & \mygreen{37}\gain{24.4} \\
Screenshot to cloud, takeover~\venue{MAI-UI form} & \underline{92.9\std{0.8}} & 6 & 2912.0 & 2.90 & \underline{539} & 9.60 & \mygreen{3}\gain{0.9} \\
Trajectory upload, cloud filter~\venue{CoMIC form} & 92.1\std{1.1} & \underline{0} & 883.9 & 0.35 & 1042 & 0.30 & \mygreen{3}\gain{1.6} \\
\midrule
\rowcolor{crystalblue!10}
\textbf{\sys{} (Ours)} & \textbf{93.8\std{1.3}} & \textbf{0} & \textbf{2.5} & \textbf{1.62} & \textbf{702} & \textbf{0.20} & -- \\
\midrule
X1: w/o update clock (Ladder only) & 90.5\std{1.0} & 0 & 2.4 & 0.37 & 1394 & 0.36 & \mygreen{5}\gain{3.3} \\
\rowcolor{crystalblue!5}
X2: w/o failure clock (Replay + Verify) & 80.6\std{1.4} & 1 & 0.0 & 1.44 & 782 & 0.00 & \mygreen{20}\gain{13.2} \\
X3: cloud blind (Verify + Ladder) & 90.5\std{2.6} & 0 & 4.1 & 0.38 & 723 & 0.39 & \mygreen{5}\gain{3.3} \\
\rowcolor{crystalblue!5}
X4: w/o device verification (Replay + Ladder) & 90.4\std{2.4} & 0 & 7.9 & 1.64 & 1403 & 0.27 & \mygreen{5}\gain{3.4} \\
X6: evidence fixed at L3 & 93.0\std{0.9} & 0 & 36.0 & 1.64 & 704 & 0.11 & \mygreen{3}\gain{0.7} \\
\rowcolor{crystalblue!5}
X7: evidence fixed at L1 & 84.0\std{1.5} & 1 & 1.0 & 1.63 & 756 & 0.11 & \mygreen{15}\gain{9.8} \\
X8: text guidance instead of patch & 80.4\std{0.8} & 1 & 1.0 & 1.52 & 766 & 0.11 & \mygreen{20}\gain{13.3} \\
\bottomrule
\end{tabular}
\end{table*}

\subsection{Experimental Setup}
\label{sec:eval-setup}
Nine version pairs, six obtainable by the cloud and three blind, each with five tasks, produce 45 evaluation units. Each method, or \emph{arm}, runs every unit for $K=5$ post-update episodes under three seeds. An episode ends upon success, after 30 steps, or after six minutes; each task is a parameterized goal whose evaluator reads the app state, as in AndroidWorld~\cite{Rawles2025ICLR}, so no model judges success. Baselines include no experience; stale experience, which executes $\Ed(v)$ on $v'$ without repair; re-explore, which records a new path after failure; and a cloud-only agent as the capability bound. Screenshot-to-cloud, a MAI-UI-style takeover baseline~\cite{Zhou2025ARXIV}, sends a screenshot after a mismatch and transfers the remaining episode to the cloud, whereas CoMIC-style trajectory upload~\cite{Wang2026ARXIVa} sends the complete episode and receives a patch. Metrics are Succ@$K$, the success rate across the $K$ episodes; the recovery index $T_r$; and uplink and downlink kilobytes, device joules, and cloud calls, the requests to the cloud model, per episode, with pushes amortized over $K$~\cite{Wu2026COMST}. We report means across units with standard deviations across seeds, and for $T_r$ the median over all unit--seed pairs. The main tier uses a $10$\,Mbps uplink and $80$\,ms round-trip time; the link-tier sweep covers $2$, $10$, and $50$\,Mbps and $20$, $80$, and $250$\,ms.

\subsection{Main Results}
\label{sec:eval-main}
Relative to stale experience, \sys{} raises Succ@$K$ by \RecoverGain{} (95\% confidence interval: \RecoverGainCI{}) and matches fresh experience recorded on the new version (difference \HMinusFresh{}; interval \HMinusFreshCI{}). Recovery occurs in the first episode for \RecoverEp{} of cloud-visible units and \RecoverEpBlind{} of blind units, compared with \RecOneRe{} under re-explore.

Screenshot-to-cloud matches \sys{} on Succ@$K$ because the cloud completes every broken episode. The device never returns to autonomous success, so $T_r$ remains censored at $K+1$. \sys{} uses \ByteSave{} fewer bytes and \CloudSave{} fewer cloud calls than this baseline, and \ByteSaveTraj{} fewer bytes than trajectory upload. Screenshot-to-cloud and cloud-only execution use less device energy (\JShot{} and \JCloudOnly{} versus \JH{}) because the cloud performs most of their inference~\cite{Duan2023COMST}.

Per episode, the device uses \JouleSave{} less energy than with no experience and \JouleSaveRe{} less than with re-explore. Verified steps require a lookup, and patched steps replay recorded actions; both reduce device-side model calls.

\subsection{Link Tiers, Fleets, and Energy}
\label{sec:eval-link}
\begin{figure*}[t]
\centering
\includegraphics[height=5.0mm]{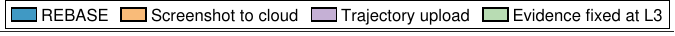}\\[0.15em]
\begin{subfigure}[t]{0.245\textwidth}\centering
\includegraphics[width=\linewidth]{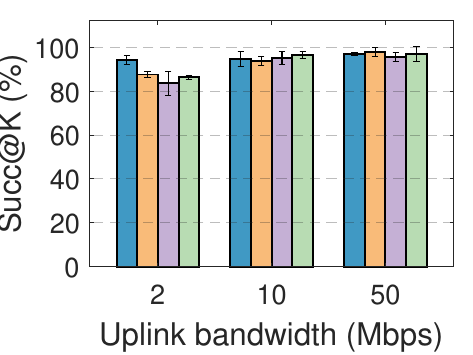}
\caption{Success vs.\ bandwidth}\label{fig:link-a}\end{subfigure}\hfill
\begin{subfigure}[t]{0.245\textwidth}\centering
\includegraphics[width=\linewidth]{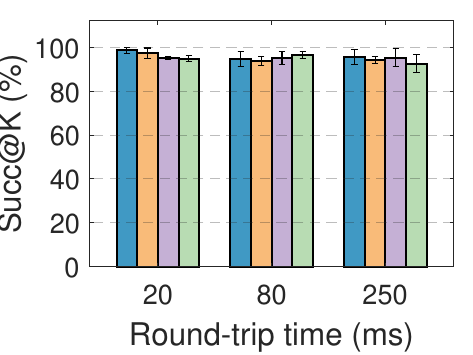}
\caption{Success vs.\ round-trip time}\label{fig:link-b}\end{subfigure}\hfill
\begin{subfigure}[t]{0.245\textwidth}\centering
\includegraphics[width=\linewidth]{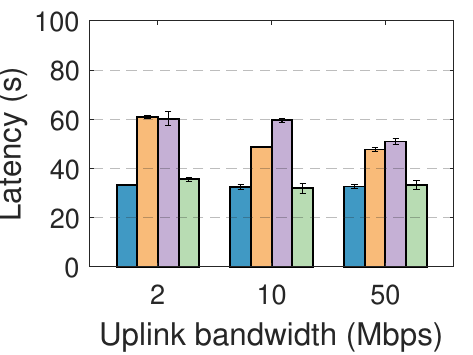}
\caption{Latency vs.\ bandwidth}\label{fig:link-c}\end{subfigure}\hfill
\begin{subfigure}[t]{0.245\textwidth}\centering
\includegraphics[width=\linewidth]{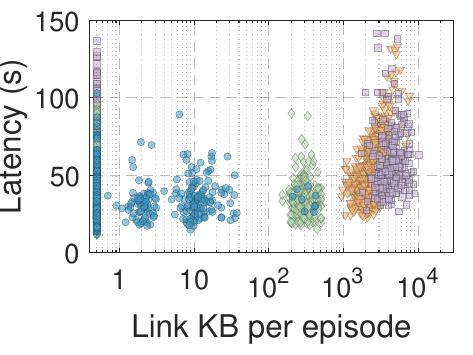}
\caption{Latency vs.\ link bytes}\label{fig:link-d}\end{subfigure}
\caption{Link tiers: three pairs, five tasks, $K=5$, and three seeds (bars: mean $\pm$ standard deviation). Succ@$K$ against (a)~uplink bandwidth at 80\,ms and (b)~round-trip time at 10\,Mbps; (c)~episode latency against bandwidth; (d)~episode latency against transferred kilobytes, with one point per episode and a 0.5\,KB floor for zero-byte episodes.}
\label{fig:link}
\end{figure*}
As bandwidth decreases from 50 to 2\,Mbps and round-trip time increases from 20 to 250\,ms, \sys{} keeps Succ@$K$ within \LinkRobustBw{} and \LinkRobustRtt{} of the fastest tier, and its episode latency stays between \LatHLo{} and \LatHHi{} (Fig.~\ref{fig:link}). The latency of screenshot-to-cloud rises from \LatShotFast{} to \LatShotSlow{} as the link slows; across arms, episode latency tracks the transferred kilobytes.

\begin{figure*}[t]
\centering
\begin{subfigure}[t]{0.245\textwidth}\centering
\includegraphics[width=\linewidth]{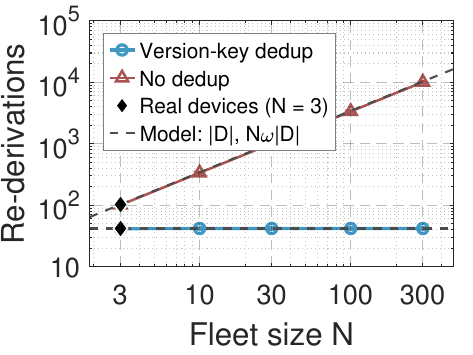}
\caption{Re-derivations vs.\ fleet size}\label{fig:fleet-a}\end{subfigure}\hfill
\begin{subfigure}[t]{0.245\textwidth}\centering
\includegraphics[width=\linewidth]{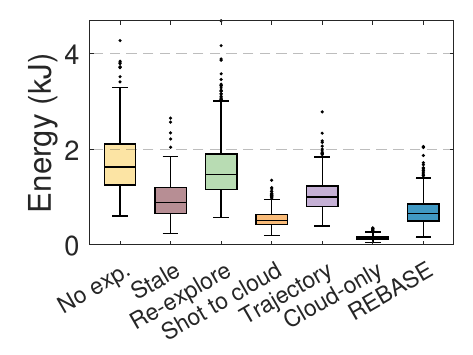}
\caption{Energy per episode}\label{fig:fleet-b}\end{subfigure}\hfill
\begin{subfigure}[t]{0.245\textwidth}\centering
\includegraphics[width=\linewidth]{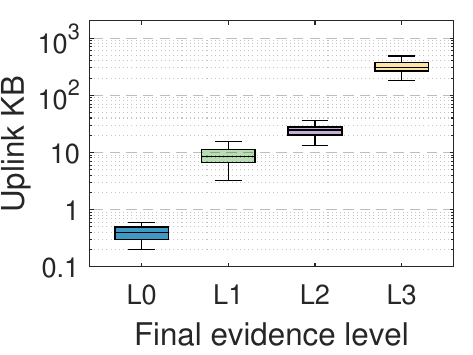}
\caption{Uplink bytes per event by level}\label{fig:fleet-c}\end{subfigure}\hfill
\begin{subfigure}[t]{0.245\textwidth}\centering
\includegraphics[width=\linewidth]{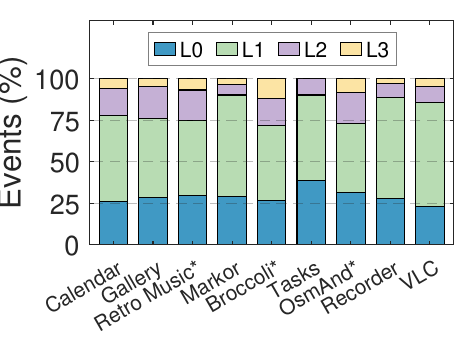}
\caption{Closing evidence level}\label{fig:fleet-d}\end{subfigure}
\caption{Fleets and energy. (a)~Cloud re-derivations vs.\ fleet size $N$ with and without the version key (trace-driven, 20 repetitions; two Orins and the Thor as real devices at $N=3$; model counts $M$ and $\tilde M$). (b)~Device energy per episode, seven arms. (c)~Uplink kilobytes per event by closing level; (d)~share of events closed per level and app, four ladder arms pooled; asterisks mark blind pairs.}
\label{fig:fleet}
\end{figure*}
With version keys, cloud re-derivations remain near $|D|=\FleetD{}$, the number of update-broken steps, as the fleet grows: in a trace-driven fleet of 100 devices they are \FleetSlope{} below the count without the key (Fig.~\ref{fig:fleet-a}); each of the three real devices encounters \FleetOverlap{} of the same broken steps.

The median event costs the device \JEventMedian{}, while each episode uses \JSavedEpisode{} less than without experience, so an event's energy is recovered within \Payback{} of an episode; \CloseByDiff{} of ladder events close at the fingerprint or diff level (Fig.~\ref{fig:fleet-c} and~\ref{fig:fleet-d}).

\subsection{Ablation Study}
\label{sec:eval-ablation}
The lower block of Table~\ref{tab:main} removes one component of \sys{} at a time. Without the update clock (X1), every run discovers the update on the device: events rise from \EventsH{} to \XoneEvents{} and cloud calls grow \XoneCloudRatio{}, stale steps per episode rise from \StaleStepsH{} to \XoneStaleSteps{}, and success falls by \DeltaXone{}. Without the failure clock (X2), success falls to \BlindNoFailClock{} from \BlindH{} on blind pairs and to \VisNoFailClock{} from \VisH{} on cloud-visible pairs. Even when the cloud cannot obtain $v'$ (X3), the failure clock keeps success within \DeltaXthree{} of \sys{}, while recovery within one episode drops from \RecOneH{} to \RecOneCloudBlind{}. Without device verification (X4), stale steps per episode rise from \StaleStepsH{} to \AprimeGain{}, and uplink traffic reaches \XfourUpRatio{} that of \sys{}. Fixing evidence at the screenshot (X6) preserves success but uses \XsixUpRatio{} the uplink bytes. Fixing it at the diff (X7) loses \DeltaXseven{} because \BeyondDiffPooled{} of ladder events need a higher level. Replacing executable patches with text guidance (X8) loses \DeltaXeight{} because the small model replays a recorded action more reliably than it interprets repair advice.

\section{Conclusion}
\label{sec:conclusion}
\sys{} maintains coherence between device and cloud copies of GUI-agent experience across app updates. The update clock combines cloud replay with device verification; the failure clock returns version-keyed patches from staged evidence. On two Jetson devices and nine Android apps, it restores stale experience to the success rate of fresh experience within one episode, with \ByteSave{} fewer bytes than screenshot upload and \JouleSave{} less device energy than no experience. Every measured failure exposed an accessibility tree; screens without one start at a crop. Desktop agents are next. Recovery depended on identifying changed elements rather than sending more pixels.

\bibliographystyle{IEEEtran}%
\bibliography{bib/refs}%

\begin{thebibliography}{10}
\providecommand{\url}[1]{#1}
\csname url@samestyle\endcsname
\providecommand{\newblock}{\relax}
\providecommand{\bibinfo}[2]{#2}
\providecommand{\BIBentrySTDinterwordspacing}{\spaceskip=0pt\relax}
\providecommand{\BIBentryALTinterwordstretchfactor}{4}
\providecommand{\BIBentryALTinterwordspacing}{\spaceskip=\fontdimen2\font plus
\BIBentryALTinterwordstretchfactor\fontdimen3\font minus
  \fontdimen4\font\relax}
\providecommand{\BIBforeignlanguage}[2]{{%
\expandafter\ifx\csname l@#1\endcsname\relax
\typeout{** WARNING: IEEEtran.bst: No hyphenation pattern has been}%
\typeout{** loaded for the language `#1'. Using the pattern for}%
\typeout{** the default language instead.}%
\else
\language=\csname l@#1\endcsname
\fi
#2}}
\providecommand{\BIBdecl}{\relax}
\BIBdecl

\bibitem{Wu2026ARXIVForget}
B.~Wu, Z.~Ding, J.~Huang, and Y.~Zhao, ``{Forget to Improve: On-Device
  LLM-Agent Continual Learning via Budget-Curated Memory},'' arXiv preprint
  arXiv:2606.25115, 2026.

\bibitem{Wu2026ARXIVERRAND}
B.~Wu, Z.~Ding, and J.~Huang, ``{ERRAND: Budgeted Maintenance of Agent
  Memory},'' arXiv preprint arXiv:2609.29545, 2026.

\bibitem{Zhou2025ARXIV}
H.~Zhou \emph{et~al.}, ``{MAI-UI Technical Report: Real-World Centric
  Foundation GUI Agents},'' arXiv preprint arXiv:2512.22047, 2025.

\bibitem{Jiang2026ACL}
Y.~Jiang and C.~Huang, ``{OpenPhone: Mobile Agentic Foundation Models},'' in
  \emph{Findings of the Association for Computational Linguistics: ACL 2026},
  2026, pp. 30\,362--30\,380.

\bibitem{Li2026ARXIV}
S.~Li, Z.~Zuo, H.~Wang, J.~Chen, Z.~Jin, and R.~LI, ``{Administrative
  Decentralization in Edge-Cloud Multi-Agent for Mobile Automation},'' arXiv
  preprint arXiv:2604.07767, 2026.

\bibitem{Fan2025NeurIPS}
G.~Fan, C.~Niu, C.~Lyu, F.~Wu, and G.~Chen, ``{CORE: Reducing UI Exposure in
  Mobile Agents via Collaboration Between Cloud and Local LLMs},'' in
  \emph{Advances in Neural Information Processing Systems}, 2025.

\bibitem{Xie2026INFOCOM}
Z.~Xie, Y.~Xu, H.~Xu, Y.~Liao, and Z.~Yao, ``{A Novel Hat-Shaped Device-Cloud
  Collaborative Inference Framework for Large Language Models},'' in
  \emph{Proceedings of the IEEE International Conference on Computer
  Communications}, 2026.

\bibitem{Wang2026INFOCOM}
Z.~Wang \emph{et~al.}, ``{PPAI: Enabling Personalized LLM Agent
  Interoperability for Collaborative Edge Intelligence},'' in \emph{Proceedings
  of the IEEE International Conference on Computer Communications}, 2026.

\bibitem{Zhang2026ARXIV}
Y.~Zhang \emph{et~al.}, ``{Hera: Learning Long-Horizon Coordination for
  Device-Cloud Collaborative LLM Agents},'' arXiv preprint arXiv:2605.24598,
  2026.

\bibitem{Fang2025JSAC}
Z.~Fang \emph{et~al.}, ``{R-ACP: Real-Time Adaptive Collaborative Perception
  Leveraging Robust Task-Oriented Communications},'' \emph{IEEE Journal on
  Selected Areas in Communications}, vol.~43, no.~12, pp. 4215--4230, 2025.

\bibitem{Qin2026ICML}
Z.~Qin, S.~Yue, X.~Hua, Y.~Fu, and J.~Ren, ``{Executable Agentic Memory for GUI
  Agent},'' in \emph{Proceedings of the International Conference on Machine
  Learning}, 2026.

\bibitem{Mi2026ICML}
H.~Mi \emph{et~al.}, ``{Darwinian Memory: A Training-Free Self-Regulating
  Memory System for GUI Agent Evolution},'' in \emph{Proceedings of the
  International Conference on Machine Learning}, 2026.

\bibitem{Men2026ACL}
T.~Men, Z.~Jin, P.~Cao, Y.~Chen, K.~Liu, and J.~Zhao, ``{Empowering GUI Agents
  via Autonomous Experience Exploration and Hindsight Experience Utilization
  for Task Planning},'' in \emph{Proceedings of the 64th Annual Meeting of the
  Association for Computational Linguistics (Volume 1: Long Papers)}, 2026, pp.
  36\,090--36\,108.

\bibitem{Wang2026ARXIVa}
Y.~Wang, L.~Yang, Z.~Liu, A.~Kumar, and C.~Maple, ``{CoMIC: Collaborative
  Memory and Insights Circulation for Long-Horizon LLM Agents in Cloud-Edge
  Systems},'' arXiv preprint arXiv:2606.00756, 2026.

\bibitem{SaadFalcon2026ARXIV}
J.~Saad-Falcon \emph{et~al.}, ``{OpenJarvis: Personal AI, On Personal
  Devices},'' arXiv preprint arXiv:2605.17172, 2026.

\bibitem{SelfHealing2026WEB}
{Healenium}, {mabl}, and {Testim}, ``{Self-healing locators in test
  automation},'' \url{https://healenium.io/docs/how_healenium_works};
  \url{https://www.mabl.com/auto-healing-tests};
  \url{https://www.testim.io/blog/announcing-auto-improving-smart-locators-dare-we-say-genius-locators/},
  accessed: 2026-09-09.

\bibitem{Cao2024ICSE}
S.~Cao \emph{et~al.}, ``{Comprehensive Semantic Repair of Obsolete GUI Test
  Scripts for Mobile Applications},'' in \emph{Proceedings of the IEEE/ACM 46th
  International Conference on Software Engineering}, 2024.

\bibitem{Cao2026PACMSE}
S.~Cao, M.~Pan, and X.~Li, ``{TUSR: A Test Unit-Based Framework for Repairing
  Obsolete GUI Test Scripts},'' \emph{Proceedings of the ACM on Software
  Engineering}, vol.~3, no. FSE, pp. 4714--4736, 2026.

\bibitem{Wu2026TNSE}
B.~Wu, Z.~Ding, and J.~Huang, ``{A Review of Continual Learning in Edge AI},''
  \emph{IEEE Transactions on Network Science and Engineering}, vol.~13, pp.
  6571--6588, 2026.

\bibitem{Wu2026ARXIVLifecycle}
B.~Wu and J.~Huang, ``{Lifecycle-Aware Federated Continual Learning in Mobile
  Autonomous Systems},'' arXiv preprint arXiv:2604.20745, 2026.

\bibitem{Android2026WEB}
{Android Developers}, ``{Create an accessibility service},''
  \url{https://developer.android.com/guide/topics/ui/accessibility/service},
  2026, accessed: 2026-09-09.

\bibitem{Wu2026ARXIVCrystalMem}
B.~Wu and J.~Huang, ``{CrystalMem: Elastic Memory for Self-Evolving LLM Agents
  via Knowledge Crystallization},'' arXiv preprint arXiv:2608.00303, 2026.

\bibitem{Fang2025ARXIV}
Z.~Fang, Y.~Guo, Y.~Zhang, H.~An, W.~Ding, and Y.~Fang, ``{Shared Spatial
  Memory Through Predictive Coding},'' arXiv preprint arXiv:2511.04235, 2025.

\bibitem{Wu2026TON}
B.~Wu, J.~Huang, Q.~Duan, L.~Dong, and Z.~Cai, ``{Enhancing Vehicular
  Platooning With Wireless Federated Learning: A Resource-Aware Control
  Framework},'' \emph{IEEE/ACM Transactions on Networking}, vol.~34, pp.
  1479--1494, 2026.

\bibitem{Dong2026TCCN}
L.~Dong, J.~Huang, and R.~W. Heath, ``{Transformer-Based Dynamic Resource
  Allocation for Multi-Carrier NOMA Systems},'' \emph{IEEE Transactions on
  Cognitive Communications and Networking}, vol.~12, pp. 4926--4941, 2026.

\bibitem{Xing2025ACR}
C.-C. Xing, Z.~Ding, and J.~Huang, ``{A Stochastic Geometry-Based Analysis of
  SWIPT-Assisted Underlaid Device-to-Device Energy Harvesting},'' \emph{ACM
  SIGAPP Applied Computing Review}, vol.~25, no.~4, pp. 18--34, 2025.

\bibitem{Wu2023ACCESS}
B.~Wu, Z.~Cai, W.~Wu, and X.~Yin, ``{AoI-Aware Resource Management for Smart
  Health via Deep Reinforcement Learning},'' \emph{IEEE Access}, vol.~11, pp.
  81\,180--81\,195, 2023.

\bibitem{Ding2026ICDCS}
Z.~Ding, B.~Wu, and J.~Huang, ``{SCALE: Sensitivity-Aware Federated Unlearning
  with Information Freshness Optimization for Mobile Edge Computing},'' in
  \emph{Proceedings of the IEEE International Conference on Distributed
  Computing Systems}, 2026, pp. 1476--1486.

\bibitem{Wu2026ICDCS}
B.~Wu, J.~Huang, and Y.~Zhao, ``{From Alpha to Omega: Lifecycle-Aware
  Forgetting Defense in Federated Continual Learning for Planetary
  Exploration},'' in \emph{Proceedings of the IEEE International Conference on
  Distributed Computing Systems}, 2026, pp. 1--11.

\bibitem{Wu2026TMM}
B.~Wu, Z.~Ding, and J.~Huang, ``{Exposing and Resolving Spurious Isolation in
  Federated Multimodal Continual Learning},'' \emph{IEEE Transactions on
  Multimedia}, 2026.

\bibitem{Wu2023MPE}
B.~Wu and W.~Wu, ``{Model-Free Cooperative Optimal Output Regulation for Linear
  Discrete-Time Multi-Agent Systems Using Reinforcement Learning},''
  \emph{Mathematical Problems in Engineering}, vol. 2023, no.~1, p. 6350647,
  2023.

\bibitem{Ding2026ICNC}
Z.~Ding, J.~Huang, and J.~Qi, ``{Learning to Defend: A Multi-Agent
  Reinforcement Learning Framework for Stackelberg Security Game in Mobile Edge
  Computing},'' in \emph{Proceedings of the International Conference on
  Computing, Networking and Communications}, 2026, pp. 769--774.

\bibitem{Huang2025RACS}
J.~Huang, B.~Wu, Z.~Ding, and L.~Ostigaard, ``{Reinforcement Learning-Based
  Energy-Aware Coverage Path Planning for Precision Agriculture},'' in
  \emph{Proceedings of the International Conference on Research in Adaptive and
  Convergent Systems}, 2025, pp. 1--8.

\bibitem{Huang2025TMC}
J.~Huang, B.~Wu, Q.~Duan, L.~Dong, and S.~Yu, ``{A Fast UAV Trajectory Planning
  Framework in RIS-Assisted Communication Systems With Accelerated Learning via
  Multithreading and Federating},'' \emph{IEEE Transactions on Mobile
  Computing}, vol.~24, no.~8, pp. 6870--6885, 2025.

\bibitem{Wu2025WASA}
B.~Wu, J.~Huang, and Q.~Duan, ``{FedTD3: An Accelerated Learning Approach for
  UAV Trajectory Planning},'' in \emph{Proceedings of the International
  Conference on Wireless Artificial Intelligent Computing Systems and
  Applications}, 2025, pp. 13--24.

\bibitem{Wu2026MNET}
B.~Wu, J.~Huang, and Q.~Duan, ``{Real-Time Intelligent Healthcare Enabled by
  Federated Digital Twins With AoI Optimization},'' \emph{IEEE Network},
  vol.~40, no.~2, pp. 184--191, 2026.

\bibitem{Ding2026ARXIVTwinLoop}
Z.~Ding, B.~Wu, J.~Huang, and S.~Mao, ``{Application-Aware Twin-in-the-Loop
  Planning for Federated Split Learning over Wireless Edge Networks},'' arXiv
  preprint arXiv:2604.26105, 2026.

\bibitem{Fang2025TON}
Z.~Fang, S.~Hu, J.~Wang, Y.~Deng, X.~Chen, and Y.~Fang, ``{Prioritized
  Information Bottleneck Theoretic Framework With Distributed Online Learning
  for Edge Video Analytics},'' \emph{IEEE/ACM Transactions on Networking},
  vol.~33, no.~3, pp. 1203--1219, 2025.

\bibitem{Ding2026ARXIVEASE}
Z.~Ding, B.~Wu, and J.~Huang, ``{EASE: Federated Multimodal Unlearning via
  Entanglement-Aware Anchor Closure},'' arXiv preprint arXiv:2605.00733, 2026.

\bibitem{Ding2026MNET}
Z.~Ding and J.~Huang, ``{Toward Trustworthy Federated Unlearning for Mobile
  Autonomous Systems},'' \emph{IEEE Network}, 2026.

\bibitem{Tchalla2026ARXIV}
D.~Y. Tchalla, B.~Wu, J.~Huang, S.~Gu, and Q.~Duan, ``{FedSceneX:
  Time-to-Target Orchestration for Same-Scene Multimodal Federated Edge
  Learning},'' arXiv preprint arXiv:2608.07730, 2026.

\bibitem{Ding2025IPCCC}
Z.~Ding, J.~Huang, Q.~Duan, C.~Zhang, Y.~Zhao, and S.~Gu, ``{A Dual-Level
  Game-Theoretic Approach for Collaborative Learning in UAV-Assisted
  Heterogeneous Vehicle Networks},'' in \emph{Proceedings of the IEEE
  International Performance, Computing, and Communications Conference}, 2025,
  pp. 1--8.

\bibitem{Wu2026IoTJ}
B.~Wu, Z.~Ding, and J.~Huang, ``{RELIEF: Turning Missing Modalities into
  Training Acceleration for Federated Learning on Heterogeneous IoT Edge},''
  \emph{IEEE Internet of Things Journal}, 2026.

\bibitem{Pan2023SCIS}
D.~Pan, B.-N. Wu, Y.-L. Sun, and Y.-P. Xu, ``{A Fault-Tolerant and
  Energy-Efficient Design of a Network Switch Based on a Quantum-Based
  Nano-Communication Technique},'' \emph{Sustainable Computing: Informatics and
  Systems}, vol.~37, p. 100827, 2023.

\bibitem{Bai2025ARXIVb}
S.~Bai \emph{et~al.}, ``{Qwen3-VL Technical Report},'' arXiv preprint
  arXiv:2511.21631, 2025.

\bibitem{Rawles2025ICLR}
C.~Rawles \emph{et~al.}, ``{AndroidWorld: A Dynamic Benchmarking Environment
  for Autonomous Agents},'' in \emph{International Conference on Learning
  Representations}, 2025.

\bibitem{Wu2026COMST}
B.~Wu, J.~Huang, and S.~Yu, ``{``X of Information'' Continuum: A Survey on
  AI-Driven Multi-Dimensional Metrics for Next-Generation Networked Systems},''
  \emph{IEEE Communications Surveys \& Tutorials}, vol.~28, pp. 5307--5344,
  2026.

\bibitem{Duan2023COMST}
Q.~Duan, J.~Huang, S.~Hu, R.~Deng, Z.~Lu, and S.~Yu, ``{Combining Federated
  Learning and Edge Computing Toward Ubiquitous Intelligence in 6G Network:
  Challenges, Recent Advances, and Future Directions},'' \emph{IEEE
  Communications Surveys \& Tutorials}, vol.~25, no.~4, pp. 2892--2950, 2023.

\end{thebibliography}

\end{document}